# Finite Element Model Updating Using Bayesian Approach

Tshilidzi Marwala[1], Lungile Mdlazi[1] and Sibusiso Sibisi[2]

[1]School of Electrical and Information Engineering
University of the Witwatersrand
Private Bag 3, WITS, 2050
South Africa
E-mail: t.marwala@ee.wits.ac.za

[2] Council for Scientific Industrial Research
P.O. Box 395, Pretoria, 0001
South Africa

**ABSTRACT**

This paper compares the Maximum-likelihood method and Bayesian method for finite element model updating. The Maximum-likelihood method was implemented using genetic algorithm while the Bayesian method was implemented using the Markov Chain Monte Carlo. These methods were tested on a simple beam and an unsymmetrical H-shaped structure. The results show that the Bayesian method gave updated finite element models that predicted more accurate modal properties than the updated finite element models obtained through the use of the Maximum-likelihood method. Furthermore, both these methods were found to require the same levels of computational loads.
*Keywords: Bayesian, Maximum-likelihood, finite element updating*

## 1. INTRODUCTION

Finite element (FE) models are widely used to predict the dynamic characteristics of aerospace structures. These models often give results that differ from measured results and therefore need to be updated to match measured results. Some of the updating techniques that have been proposed to date use time, modal, frequency and time-frequency domain data [1, 2]. In this paper, modal domain data is used to update the FE model. A literature review on FE updating [1] reveals that the updating problem has been mainly framed in the Maximum-likelihood framework. Even though this framework has been applied successfully in industry, it has the following shortcomings: it does not offer the user confidence intervals for solutions it gives; there is no philosophical explanation of the regularization terms that are used to control the complexity of the updated model; and it cannot handle inherent ill-conditioning and non-uniqueness of FE updating problem. The Bayesian framework is adopted to address the shortcomings explained above. Bayesian framework has been found to offer several advantages over Maximum-likelihood methods in areas closely mirroring FE updating [3-5]. This paper seeks to address the following issues: (1) how prior information are incorporated into FE model updating problem; and (2) applying Bayesian framework to update FE models to match experimentally measured modal properties (i.e. natural frequencies and mode shapes) to modal properties calculated from the FE model of a beam. In this Paper Markov chain Monte Carlo (MCMC) simulation [3] is used to sample the probability of the updating parameters in light of the measured modal properties. This probability is known as the posterior probability. Metropolis algorithm [6] is used as an acceptance criterion when sampling the posterior probability.

## II. MATHEMATICAL FOUNDATION

### A. Dynamics

All elastic structures may be described in terms of their distributed mass, damping and stiffness matrices. If damping terms are neglected, the dynamic equation may be written in modal domain (natural frequencies and mode shapes) for the $i^{th}$ mode as follows [7]:

$$(-\omega_i^2[M]+[K])\{\phi\}_i = \{\varepsilon\}_i \quad (1)$$

Here $[M]$ is the mass matrix, $[K]$ is the stiffness matrix, $\omega_i$ is the $i^{th}$ natural frequency, $\{\phi\}_i$ is the $i^{th}$ mode shape vector and $\{\varepsilon\}_i$ is the $i^{th}$ error vector. The error vector $\{\varepsilon\}_i$ is equal to $\{0\}$ if the system matrices $[M]$ and $[K]$ correspond to the modal properties. If the system matrices which are usually obtained from the FE model do not match measured modal properties $\omega_i$ and $\{\phi\}_i$ then $\{\varepsilon\}_i$ is a non-zero vector. In maximum-likelihood method the Euclidean norm of $\{\varepsilon\}_i$ is minimized in order to match the system matrices to measured modal properties. Another problem that is encountered in many practical situations is that the dimension of mode shapes does not match the dimension of system matrices. This is because measured modal co-ordinates are fewer than FE modal co-ordinates. To ensure compatibility between system matrices and mode shape vectors, the dimension of system matrices is reduced by using a technique called Guyan reduction method [8] to match the dimension of system matrices to the dimension of measured mode shape co-ordinates.

### B. Bayesian Method

In this paper the Bayesian method is introduced to solve the FE updating problem based on modal properties. The fundamental rule that governs the Bayesian approach is written as follows [3]:

$$P(\{E\}|[D]) = \frac{P([D]|\{E\})P(\{E\})}{P([D])} \quad (2)$$

Here $\{E\}$ is a vector of updating parameters, $P(\{E\})$ is the probability distribution function of updating parameters in the absence of any data, and this is known as the prior distribution, and $[D]$ is a matrix containing natural frequencies $\omega_i$ and mode shapes $\{\phi\}_i$. It must be d that the mass $[M]$ and stiffness $[K]$ matrices are functions of updating parameters $\{E\}$. The quantity $P(\{E\}|[D])$ is the posterior distribution function after a set of data has been seen, $P([D]|\{E\})$ is the likelihood distribution function and $P([D])$ is the normalization factor.

*Likelihood distribution function*
There are many areas where the likelihood distribution function has been applied and these include neural networks [3]. In neural network context, the likelihood distribution function is defined as the normalized exponent of the error function. In this paper the likelihood distribution function, $P([D]|\{E\})$, is defined as the sum of square of elements of the error vector shown in equation 1, and can be written in the same way as in neural networks as follows [3]:

$$P([D]|\{E\}) = \frac{1}{Z_D(\beta)} \exp\left(-\beta \sum_j^F \sum_i^N \left(\varepsilon_{ij}^2\right)\right)$$
$$= \frac{1}{Z_D(\beta)} \exp\left(-\beta \sum_j^F \sum_i^N \left[\left(-\omega_i^2[M]+[K]\right)\{\phi\}_i\right]_j^2\right) \quad (3)$$

Here $\beta$ is the coefficient of the measured modal property data contribution to the error and is set to 1 through trial and error and $\varepsilon_{ij}$ is the error matrix with subscript i representing the $i^{th}$ modal properties and j representing the $i^{th}$ measurement position. The superscript F is the number of measured mode shape coordinates, N is the number of measured modes and ZD is:

$$Z_D(\beta) = \int \exp\left(-\beta \sum_j^F \sum_i^N \left([(-\omega_i^2[M]+[K])\{\phi\}_i]_j\right)^2\right) d[D] \qquad (4)$$

It should be d that in equation 3 the error $\varepsilon_{ij}$ is a matrix as opposed to a vector as is the case in equation 1. This is because it takes into account of all modal co-ordinates.

*Prior distribution function of parameters to be updated*
The prior distribution function consists of the information that is known about the problem. In FE updating it is generally accepted that FE updating is usually valid if the model is close to the true model. In this paper, it is known that not all parameters to be updated have the same level of modeling errors. This means that some parameters are to be updated more intensely than others. For example, parameters next to joints should be updated more intensely than those with smooth surface areas and are far from joints. In this Paper the prior distribution function for parameters to be updated may be written by using Gaussian assumption as follows [3]:

$$P(\{E\}) = \frac{1}{Z_E(\alpha)} \exp\left(-\sum_i^Q \frac{\alpha_i}{2}\|\{E\}\|^2\right) \qquad (5)$$

Here Q is the number of groups of parameters to be updated, $\alpha_i$ is the coefficient of the prior distribution function for the $i^{th}$ group of updating parameters. The prior distribution function in equation 5 ensures that large updating of parameters is less likely than small adjustments of updating parameters. The Gaussian prior has been successfully used [3] to identify a large number of weights in neural networks, and therefore it is assumed that it should be successful on identifying a small number of updating parameters in this Paper. The higher the $\alpha_i$ the lower is the degree of updating of the $i^{th}$ group of parameters and $\|\bullet\|$ is the Euclidean norm of $\bullet$. In equation 5, if $\alpha_i$ is constant for all the updating parameters, then the updated parameters will be of the same order of magnitudes. Equation 5 may be viewed as a regularization parameter [9]. In equation 5, Gaussian priors are conveniently chosen because many natural processes tend to have Gaussian distribution. In Bayesian framework regularization method is viewed as a mechanism of incorporating prior information whereas in maximum-likelihood method they are viewed as mathematical convenience. The function $Z_E(\alpha)$ is a normalization factor given by [3]:

$$Z_E(\alpha) = \int \exp\left(-\sum_i^Q \frac{\alpha_i}{2}\|\{E\}\|^2\right) d\alpha_i \qquad (6)$$

*Posterior distribution function of weight vector*
The distribution of the weights $P(\{E\}|[D])$ after the data have been seen is calculated by substituting equations 3 and 5 into equation 2 to give:

$$P(\{E\}|[D]) = \frac{1}{Z_s(\alpha,\beta)} \exp\left(-\beta \sum_j^F \sum_i^N \left([-\omega_i^2[M]+[K])\{\phi\}_i]_j\right)^2 - \sum_i^Q \frac{\alpha_i}{2}\|\{E\}\|^2\right) \qquad (7)$$

where

$$Z_S(\alpha,\beta) = P([D]) = \int \exp\left(-\beta \sum_j^F \sum_i^N \left([(-\omega_i^2[M]+[K])\{\phi\}_i]_j\right)^2 - \sum_i^Q \frac{\alpha_i}{2}\|\{E\}\|^2\right) d\{E\} \qquad (8)$$

In equation 7, the optimal weight vector corresponds to the maximum of the posterior distribution function, which is the solution as the one obtained from a maximum-likelihood approach. This implies that Bayesian method at least gives the solution that is given by the Maximum-likelihood method but in addition gives probability distributions.

**C. Markov chain Monte Carlo method**

The application of Bayesian approach to FE model updating using Monte Carlo approach, results with a set of updated parameter vectors $\{E\}_i$ that are statistical rather than deterministic. As a result, the FE model updating will give distributions of the predicted modal properties and from these distributions averages and variances of modal properties may be constructed. Following the rules of probability theory, the distribution of vector $\{y\}$, representing measured modal properties may be written in the following form:

$$P(\{y\}|[D]) = \int P(\{y\}|[E])P(\{E\}|[D])d\{E\} \quad (9)$$

Equation 9 depends on equation 7, and is difficult to solve analytically due to the relatively high dimension of updating parameter vector. As a result, Markov Chain Monte Carlo (MCMC) method is employed to determine the distribution of updating parameters, and subsequently, the distribution of predicted modal properties. The integral in equation 9 is solved, using Metropolis algorithm [6], through generating a sequence of vectors $\{E\}_1$, $\{E\}_2$,…that form a Markov chain with a stationary distribution $P([D]|\{E\})$. The integral in equation 9 may be thus approximated as follows:

$$\{\tilde{y}\} \cong \frac{1}{L}\sum_{i=I}^{R+L-1} G(\{E\}_i) \quad (10)$$

Here G is a finite element model which takes vector $\{E\}_i$ and predicts the average output $\{y\}$, which is the vector containing the modal properties, R is the number of initial states that are discarded in the hope of reaching a stationary distribution described by equation 7 and L is the number of retained states. Several methods have been proposed to simulate the distribution in equation 7 such as Gibbs sampling [10], Metropolis algorithm [6] and hybrid Monte Carlo method [11]. Hybrid Monte Carlo, which has been shown to be the most efficient of the Monte Carlo methods thus far, is not used here because it requires gradient information, which is not available in exact form in FE updating problem. As a result, MCMC method is used to identify the posterior distribution function of the updating parameters. In this Paper MCMC method is implemented by sampling a stochastic process consisting of random variables $\{\{E\}_1,\{E\}_2,…,\{E\}_n\}$ through introducing random changes to updating parameter vector $\{E\}$ and either accepting or rejecting the sample according to Metropolis algorithm [6]. Metropolis criteria can be written as follows:

$$\text{if } P_{new}(\{E\}|[D]) > P_{old}(\{E\}|[D]) \text{ accept state } \{E\}_{new}$$
$$\text{else accept } \{E\}_{new} \text{ with probability } \frac{P_{new}(\{E\}|[D])}{P_{old}(\{E\}|[D])} \quad (11)$$

In this Paper we view this procedure as a way of generating a Markov chain with transition from one state to another conducted using the criterion in equation 11.

### III. EXAMPLE: EXPERIMENTALLY MEASURED BEAM

To test the proposed procedure a freely suspended aluminum beam is used. The beam, which is shown in Figure 1, has the following dimensions: length: 1.0m; width: 25.4mm and thickness: 13.4mm. Acceleration measurements are taken at 13 equidistant positions and the beam is excited at a position located 420mm from the end of the beam (see Figure 1).

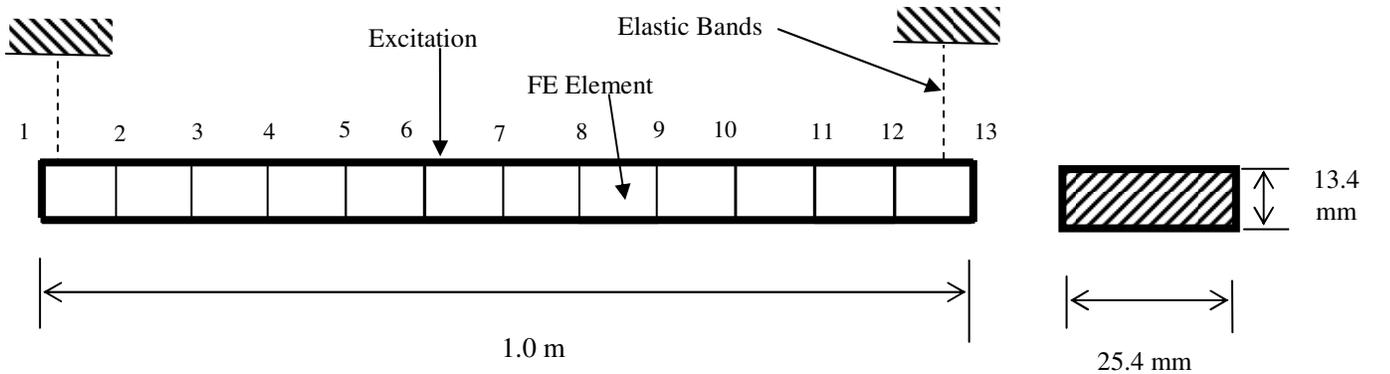

*Figure 1: A diagram showing a beam showing, its cross-sectional area, elastic bands used for suspension and positions where acceleration measurements were taken*

Further details of this beam are shown by Marwala [12]. The FE model with 12 elements is constructed using modulus of elasticity of 800×108Nm-2 and density of 2700 kgm-2. Using conventional signal processing analysis [7], the measured data are transformed into frequency response functions (FRFs) and from the FRFs, natural frequencies and mode shapes are extracted using modal extraction techniques [7]. Using the extracted natural frequencies and mode shapes the FE model is updated using Bayesian framework. When applying Bayesian framework equation 7 is used and prior information is divided into four parts, each with its own coefficient of prior distribution ($\alpha_1$, $\alpha_2$, $\alpha_3$ and $\alpha_4$). These coefficients are also shown in equation 7 by setting Q equals to 4. The coefficient $\alpha_1$ is associated with the density of the beam and is known to be uniform for all elements and is also known to be fairly accurate. The coefficient $\alpha_1$ is set to 10 to ensure that the density of the beam is not updated significantly. The coefficient $\alpha_2$ is associated with the moduli of elasticity of all elements. All elements are known to have uniform modulus of elasticity which is known fairly accurately. The coefficient $\alpha_2$ is set to 10 to ensure that the modulus of elasticity is not updated significantly. The coefficient $\alpha_3$ is associated with the cross-sectional areas of elements 1-4 and 7-12, which are known fairly accurately. The coefficient $\alpha_3$ is set to 10 to ensure that cross-sectional areas of these elements are not updated significantly. The coefficient $\alpha_4$ is associated with cross-sectional areas of elements 5 and 6, which are not accurately known because they enclose the area which was drilled to mount the excitation device. The coefficient $\alpha_4$ is set to 0.1 to ensure that cross-sectional areas of these elements are updated significantly. The MCMC method is implemented by employing the Metropolis acceptance criterion (see equation 11) and 1000 samples are retained to form a posterior distribution function indicated by equation 7.

## IV. DISCUSSION

When natural frequencies from the updated FE model are compared to those calculated from the initial FE model as well as those from the measured natural frequency data, the results in Table 1 are obtained. Table 1 also shows standard deviations of the distributions obtained through the use of the MCMC method to sample distribution in equation 7.

Table 1: Measured natural frequencies, those calculated from an initial FE model, updated FE model and associated standard deviations Key: FE: finite element

| Mode Number | Experiment (Hz) | Initial FEM (Hz) | Average Updated (Hz) | Standard Deviation (Hz) |
| --- | --- | --- | --- | --- |
| 1 | 64 | 70 | 67 | 2.8 |
| 2 | 184 | 193 | 183 | 7.6 |
| 3 | 349 | 379 | 360 | 16.1 |
| 4 | 599 | 628 | 590 | 28.7 |
| 5 | 898 | 942 | 893 | 76.4 |

The updated natural frequencies are calculated using equation 10. This table shows that for all the modes the updated model is more accurate than the initial model. Furthermore, it is observed in Table 1 that the higher the mode the higher is the standard deviation indicating that higher modes are less certain than lower modes. This is consistent with the knowledge that in general high frequency modes are less certain than low frequency modes. To compare analytical mode shapes to measured mode shapes, the modal assurance criterion (MAC) is used [13]. The MAC is a criterion that represents how well two mode shapes are correlated. Two perfectly correlated mode shapes give an identity matrix. As a result, in this paper the diagonals of the MAC whose elements are supposed to be equal to 1 for similar mode shapes are used to assess the effectiveness of the proposed updating method. The diagonal of the MAC between mode shapes from experiment and those from the updated FE models are shown inTable 2.

Table 2. Modal assurance criterion between the measured mode shapes and FE model calculated mode shapes as well as associated standard deviations. Key: MAC: modal assurance criterion.

| Mode number | MAC Experiment/initial | Average MAC Experiment/ Updated | Standard Deviation |
|---|---|---|---|
| 1 | 0.9961 | 0.9992 | 0.0011 |
| 2 | 0.9895 | 0.9974 | 0.0019 |
| 3 | 0.9799 | 0.9958 | 0.0029 |
| 4 | 0.9703 | 0.9981 | 0.0011 |
| 5 | 0.9712 | 0.9949 | 0.0044 |

This table shows that the updated FE model produces more accurate mode shapes than the initial FE model. Tables 1 and 2 show the standard deviations are used to construct error bars that measure confidence intervals of updated models. Figure 2shows the distributions of the first natural frequency and mode shape co-ordinate. . From these distributions error bars may be constructed for confidence intervals.

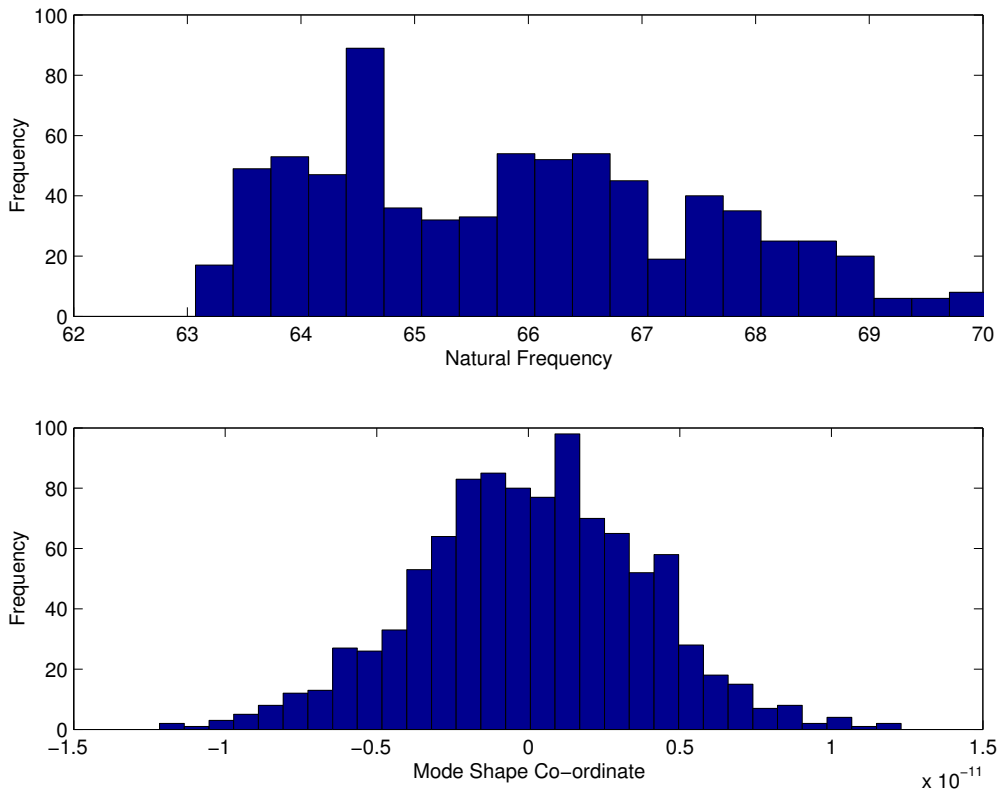

Figure 2: Sample distributions of the natural frequencies and mode shape co-ordinates of the updated FE model prediction. Key: FE: finite element

### V.  CONCLUSIONS

In this paper an updating procedure, which uses Bayesian framework and modal properties is implemented using Markov Chain Monte Carlo method. The method takes prior information into account and has an advantage of giving distributions of predicted modal properties. When the method is tested on experimental data it is found to significantly improve the accuracy of finite element models.